\documentclass{article}

\usepackage{epsfig}
\usepackage{graphicx}

\begin{document}
%\date{\today}
%\pagestyle{plain}
%% uncomment the following line to get equations numbered by (sec.num)
%\eqsec
%\newcount\eLiNe\eLiNe=\inputlineno\advance\eLiNe by -1
\title{Comparison study of DFA and DMA methods in analysis of autocorrelations in time series}
%\thanks{ {\tt grech@ift.uni.wroc.pl}}%

\author{Dariusz GRECH
\\
Institute of Theoretical Physics, Wroclaw University, pl. M. Borna
9, \\ 50-405 Wroc{\l}aw, Poland\thanks{\tt dgrech@ift.uni.wroc.pl}
 \\ and \\
  Zygmunt MAZUR
\\
Institute of Experimental  Physics, Wroclaw University, pl. M.
Borna 9,\\ 50-405 Wroc{\l}aw, Poland\thanks{\tt zmazur@ifd.uni.wroc.pl}}

\date{}
 \maketitle

\begin{abstract}
Statistics of the Hurst scaling exponents calculated with the use
of two methods: recently introduced Detrended Moving Average
Analysis(DMA) and Detrended Fluctuation Analysis (DFA)are
compared. Analysis is done for artificial stochastic Brownian time
series of various length and reveals interesting  statistical
relationships between two methods. Good agreement between DFA and
DMA techniques is found for long time series $L\sim 10^{5}$,
however for shorter series we observe  that two methods give
different results with no systematic relation between them. It is
shown that, on the average, DMA method overestimates  the Hurst
exponent comparing it with DFA technique.
\end{abstract}

\section{Introduction}
The main problem discussed in the context of stochastic time
series in various physical, biological, financial and economical
processes is the presence of autocorrelations in data. One of the
technique to check whether such autocorrelations are present in
time series is based on the investigation of the fractal structure
in time series and is related to the scaling exponent H, sometimes
denoted also as $\alpha$ \cite{bgr1}--\cite{bgr3} and called Hurst
exponent. It plays a significant role as the main concept upon
which fluctuations of a time series around its local trend (drift)
are formed and it may be considered as the one of the crucial
points responsible for 'genetic code' of time series of various
origin. For the purpose of mentioned above fractal analysis one
can introduce the scaling exponent $\alpha$ as follows.

 Let $x(t)$
($t=1,...,L$) is the time series defined for discrete time
points~$t$. By rescaling time axis $\gamma$ times (e.g. enlarging
it $\times 10^{n}$), one reveals the tiny structure of time series
not visible for smaller resolution ($\gamma \sim 1$). The fractal
structure of the series comes from the relation:

\begin{eqnarray}
\label{gr1} x'(t')\equiv\Gamma x(\gamma^{-1}t)\sim x(t)
\end{eqnarray}
where $\sim $ means similarity correspondence.

The above formula indicates that the magnitude of rescaled  time
series $x(\gamma^{-1}t)$ should be simultaneously increased
$\Gamma$ times in order to satisfy full (local) equivalence of
$x(t)$ and $x'(t')$ series.
\\
It turns out that the scaling factor $\Gamma $ can be expressed in
terms of time rescaling factor $\gamma $ with the use of
Hurst-Hausdorff $\alpha $ exponent ($\alpha > 0 $):

\begin{eqnarray}
\label{gr2}
\Gamma = \gamma^\alpha
\end{eqnarray}

%WSTAWKA!!!
The commonly accepted methods to measure $\alpha$ exponent are
Rescaled Range Analysis (R/S), spectral density analysis
\cite{bgr15}, and Detrended Fluctuation Analysis (DFA)
\cite{bgr4}.
 %% \cite{grma1,grma7} \cite{grm1}--\cite{grm14}.
 Recently, new method called Detrended Moving
Average (DMA) has also been proposed \cite{bgr5,bgr6}. In this
article we will focus on the latter two methods due to large
uncertainties in spectral density analysis and problems with R/S
predictions in nonstationary series.
%%%%%
 Searches for better
understanding how results of these two methods relate to each
other are in progress \cite{bgr6}--\cite{bgr8}.
%%%%%
A DFA method was first developed for biological purposes
\cite{bgr4} and then applied also to finances
\cite{bgr9}--\cite{bgr11}. It is a detrendisation technique
basically measuring fluctuations of a given time series around its
local trend as a function of the trend length. Let us recall the
main steps of this method:

\begin{enumerate}

\item A given signal $x(t)$ ($t=1,...,L$) of time series is
divided into $L/\tau$ not overlapping boxes of length $\tau$ each.

\item A polynomial fit $x_{\tau,k}$ is constructed in each box
representing the local trend in that box, where $k$ is the order
of polynomial fit. \item A detrended signal $X_{\tau,k}(t)$ is
found:
\begin{eqnarray}
\label{gr3}X_{\tau,k}(t) = x(t) - x_{\tau,k}(t)
\end{eqnarray}
and then its fluctuation (standard deviation)$F_{DFA}(\tau,k)$  is
calculated
\begin{eqnarray}
\label{gr4}
F_{DFA}(\tau,k) = \left({\frac{1}{L}\sum_{t=1}^L
X^2_{\tau,k}(t)}\right)^{1/2}
\end{eqnarray}

\item From the basic differential stochastic equation of the time
series $x(t)$ with a local drift $\mu(t)$ and a local dispersion
$\sigma(t)$

\begin{eqnarray}
\label{gr5}dx(t) = \mu(t)dt + \sigma(t)dX(t)
\end{eqnarray}
one expects the power law behavior:
\begin{eqnarray}
\label{gr6}F_{DFA}(\tau,k) \sim \tau^{\alpha(k)}
\end{eqnarray}
where $\alpha(k)$ is the searched Hurst exponent.

\end{enumerate}

The last equation enables to calculate $\alpha$ exponent directly
from log-log linear fit:

\begin{eqnarray}
\label{gr7}\log F_{DFA}(\tau,k) \sim \alpha(k)\log \tau
\end{eqnarray}
It can be proved that $\alpha(k)$ depends very weakly on $k$
\cite{bgr11,bgr12} so in most application one takes linear
function $(k=1)$ as a good candidate for $x_{\tau,k}$. This
approach will also be used in our paper.

 It turns out that the bigger $\alpha$ the more 'quiet' time series
is, i.e. a signal fluctuates in a more correlated way. In fact,
for $0<\alpha<1/2$ we have negative autocorrelations
(antipersistence) in time series. On the other hand, if
$1/2<\alpha\leq 1$, there are positive autocorrelations
(persistence) in signal. The case $\alpha = 1/2$ corresponds to
completely uncorrelated signal, so called integer Brownian walk.
An existing link between $\alpha$ exponent and the probability
that a given trend will last in the immediate future if it did so
in the immediate past gives an additional hint about trend changes
forecast possibility \cite{bgr13}.

A Detrended Moving Average (DMA) technique looks very similar to
DFA. The main difference one meets here is that instead of linear
or polynomial detrendisation procedure in equally sized boxes, one
uses moving average of a given length $\lambda$. The basic steps
of DMA analysis are then:

\begin{enumerate}
\item A simple moving average of length $\lambda$ ($\lambda =
1,...,L$) is constructed for $x(t)$ series $(t\geq \lambda)$:
\begin{eqnarray}
\label{gr8}\langle x(t)\rangle_\lambda =
\frac{1}{\lambda}\sum_{k=0}^{\lambda-1} x(t-k)
\end{eqnarray}
\item A detrended signal is  found similarly to Eq.~(\ref{gr3}):
\begin{eqnarray}
\label{gr9}X_{\lambda}(t) = x(t) - \langle x(t)\rangle_\lambda
\end{eqnarray}
and its fluctuation within a window of size $\lambda$ reads now:
\begin{eqnarray}
\label{gr10}F_{DMA}(\lambda) = \left({\frac{1}{L-\lambda
+1}\sum_{t=\lambda}^L X^2_\lambda(t) }\right)^{1/2}
\end{eqnarray}
\item Similarly to DFA a power law should be observed
\begin{eqnarray}
\label{gr11} \log F_{DMA}(\lambda) \sim \alpha \log\lambda
\end{eqnarray}
where $\alpha$ is the searched scaling Hurst exponent.
\end{enumerate}

The DMA technique is less complicated and seems to be faster in
practical application than DFA algorithm. However, so far no final
clear conclusion has been reached regarding mutual relationship
between DFA and DMA results for the same series.This article
contributes to the above area of interest.

\section{DMA--DFA Comparison Study}

Preliminary results obtained for some real financial series
\cite{bgr6} suggest that $\alpha_{DMA}$ values are lower than
corresponding $\alpha_{DFA}$ results. It seems to be confirmed for
the set of artificial time series of length $L\sim 2^{18}$
constructed with the use of Random Midpoint Displacement (RMD)
algorithm where one finds $\alpha_{DFA}\sim\alpha_{DMA}+0.05$
\cite{bgr5}. This supports the existence of systematic
displacement between DFA and DMA results, at least for longer
series. In many practical applications however, the length of time
series we deal with is shorter (e.g. finance, biology, genetics,
medicine), especially if one looks at the \textit{local $\alpha$
exponent} value rather than the global one \cite{bgr9}.

To attack the problem of mutual dependence between DMA and DFA
results for series of various length, let us first look at the set
of artificial arithmetic
 integer Brownian time series of length $L=3\times 10^4$ with discrete
time interval $\Delta t = 1$, i.e.:
\begin{eqnarray}
\label{gr12}
 x(L \Delta t) = x_0 + \sum_{k=1}^L \Delta x_k
\end{eqnarray}
where $\Delta x_k$ ($k=1,...,L$) are centered and normalized
displacements generated by random number generator.

Two cases with opposite relation $\alpha_{DMA}$ vs $\alpha_{DFA}$
are shown in {Fig.~1}. In the first case
$\alpha_{DFA}>\alpha_{DMA}$ and $\alpha_{DFA} - \alpha_{DMA} =
0.02$, in the other one $\alpha_{DFA}<\alpha_{DMA}$ and
$\alpha_{DMA}-\alpha_{DFA} = 0.04$. Thus no systematic
relationship is produced.

\begin{figure}[hb]
\begin{center}
{\epsfig{file=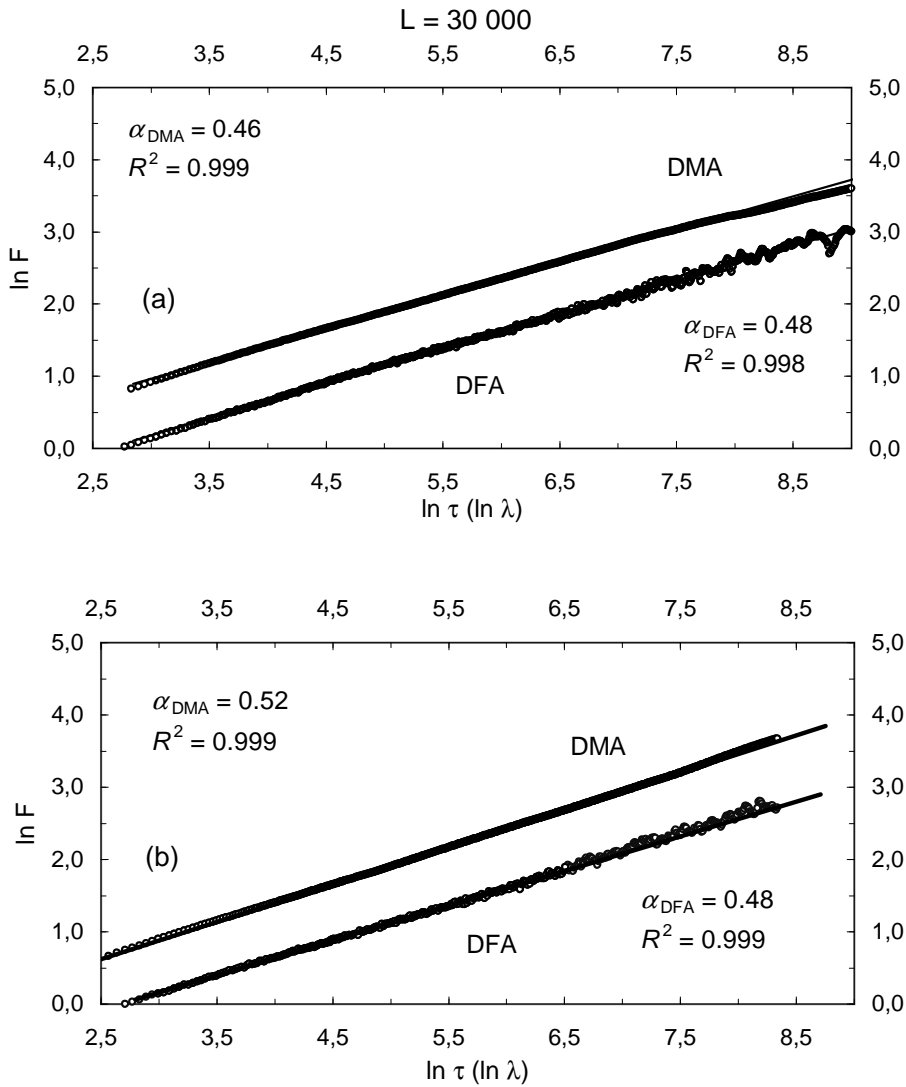,width=0.7\textwidth,angle=0}}
\caption{
Examples of DFA and DMA $\alpha$ exponent fit for artificial
Brownian
  time series of length $L=30 000$, where (a)$\alpha_{DFA}>\alpha_{DMA}$
  and (b)$\alpha_{DFA}<\alpha_{DMA}$}
\end{center}
\end{figure}

%%%%%

This induces to treat the problem statistically, i.e. one should
find statistical distributions of Hurst exponents measured within
two methods for artificial series of various length. It seems to
be interesting to compare two statistics and to work out
correlations between scaling exponents measured within DMA and DFA
techniques for the same sample of time series.

%%%%%%%%%%%%%%%%%%%
\begin{figure}
\begin{center}
{\epsfig{file=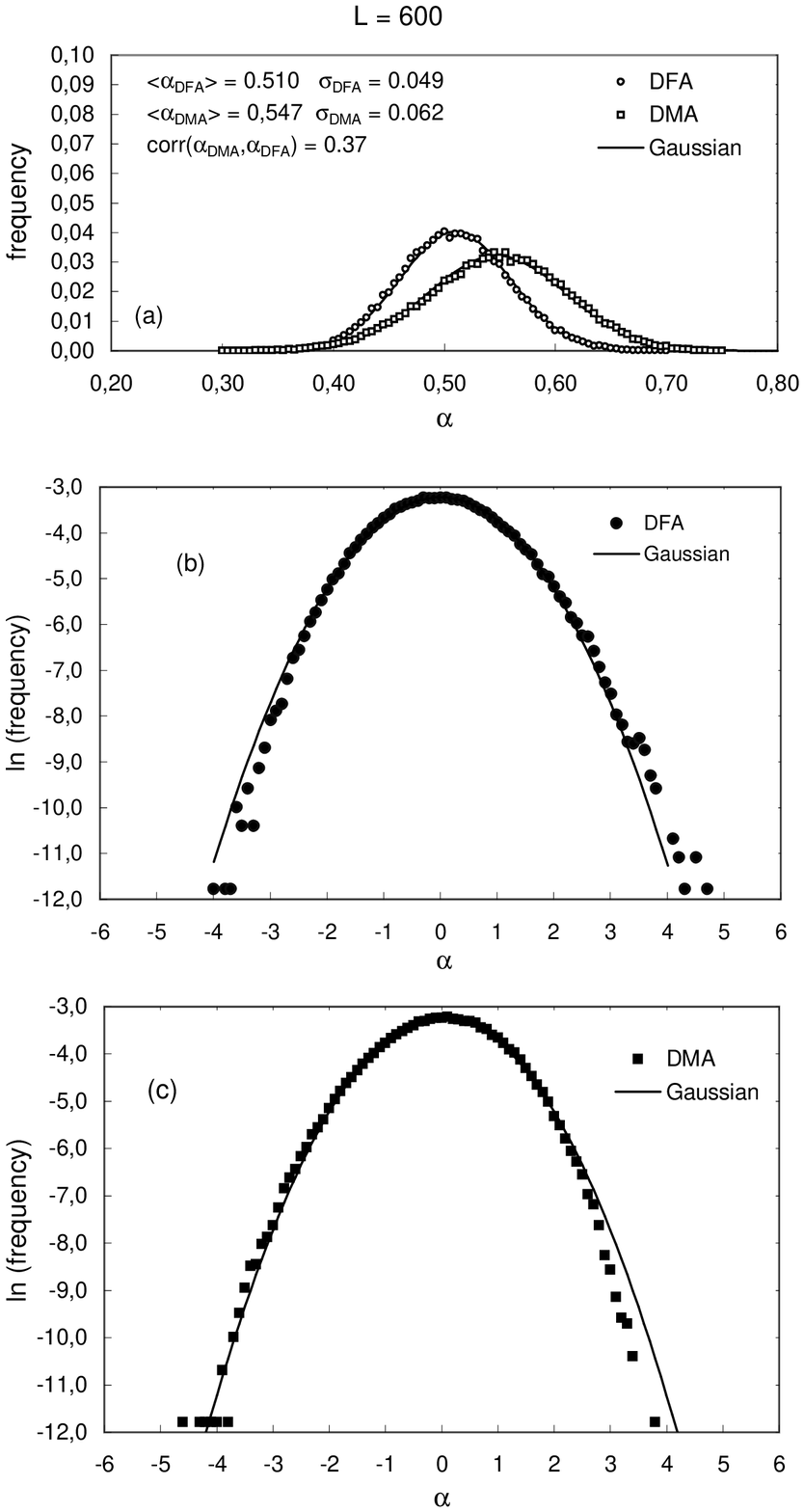,width=0.7\textwidth,angle=0}}
%\\
\caption{ (a)Distribution of scaling $\alpha$ exponent obtained
with the use of DFA (circles)
  and DMA (squares) techniques for the sample of $65 000$ series of length $L=600$.
  The normal distribution fit with corresponding parameters is also shown as a solid line.
  (b)(c) The same plots for DFA(b) and DMA(c)in semi-log scale  for normalized and centered $\alpha$ exponents. }
\end{center}
\end{figure}

%%%%%% ??????????????????????????????
%%%
\begin{figure}
\begin{center}
{\epsfig{file=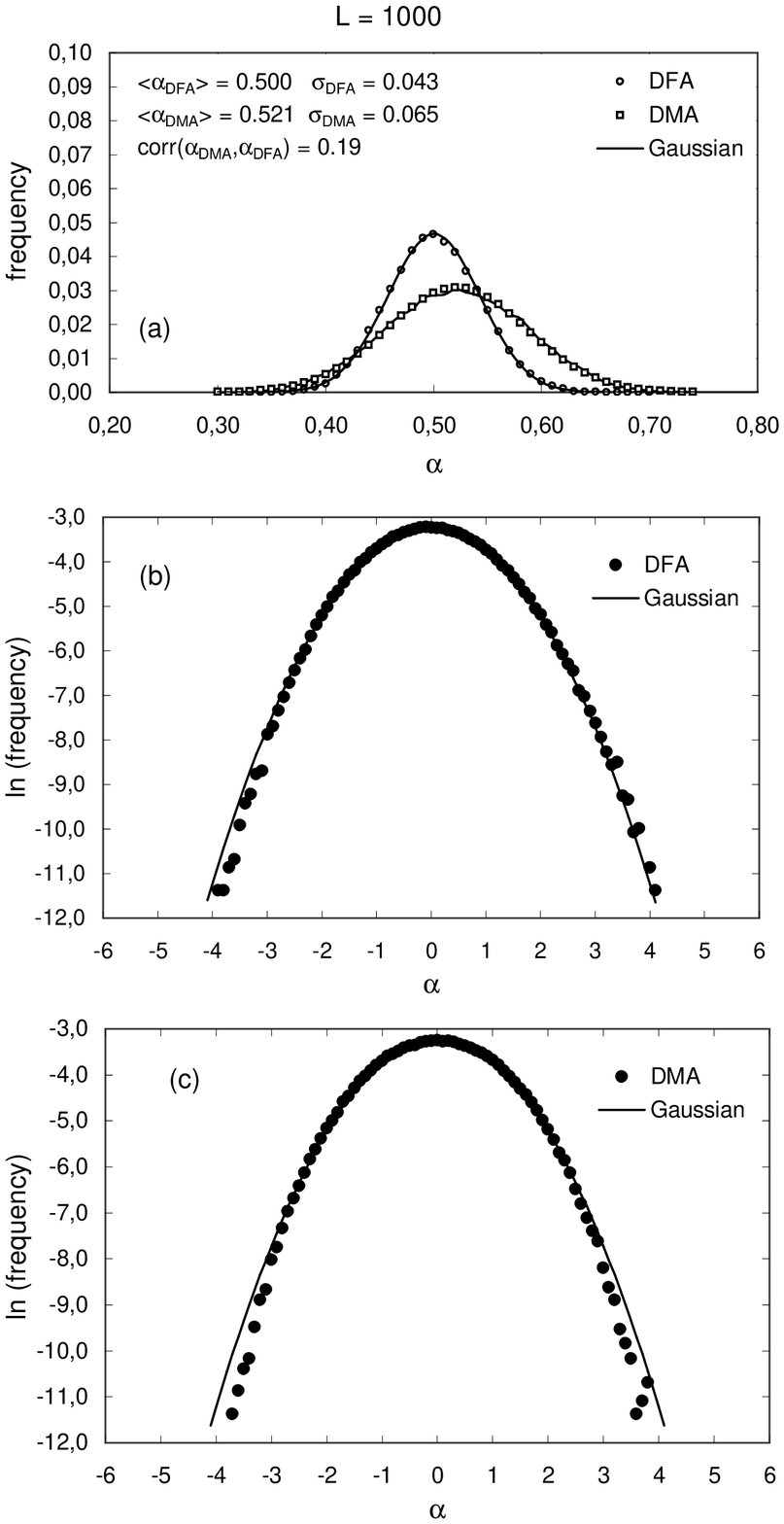,width=0.7\textwidth,angle=0}}
%\\
\caption{(a)Distribution of $\alpha$ exponents for series with
$L=1000$.
  (b)(c) Corresponding plots in semi-log scale.  }
\end{center}
\end{figure}
\begin{figure}
\begin{center}
{\epsfig{file=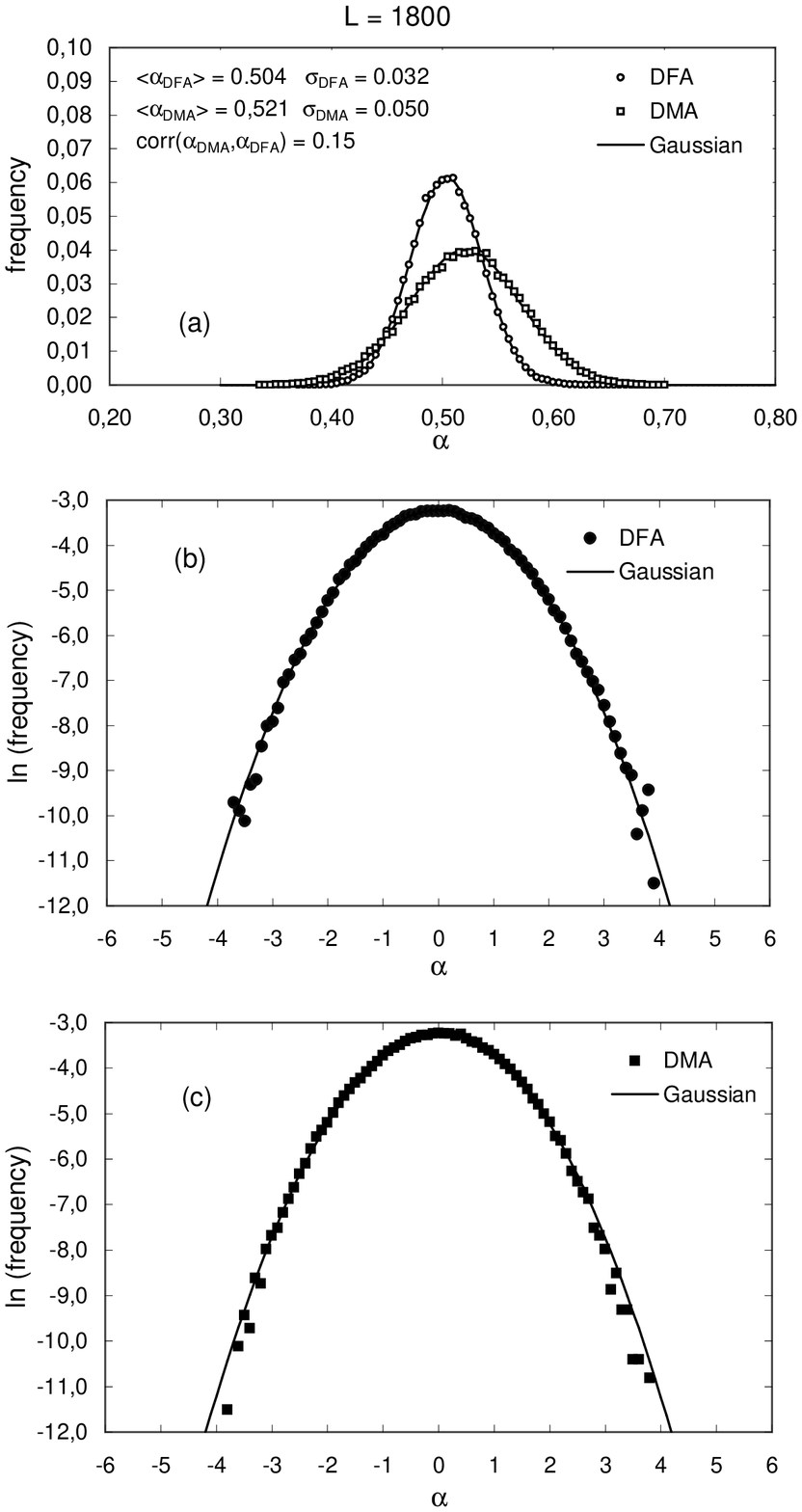,width=0.7\textwidth,angle=0}}
%\\
\caption{(a)Distribution of $\alpha$ exponents for series with
$L=1800$.
  (b)(c) Corresponding plots in semi-log scale.  }
\end{center}
\end{figure}

\begin{figure}
\begin{center}
{\epsfig{file=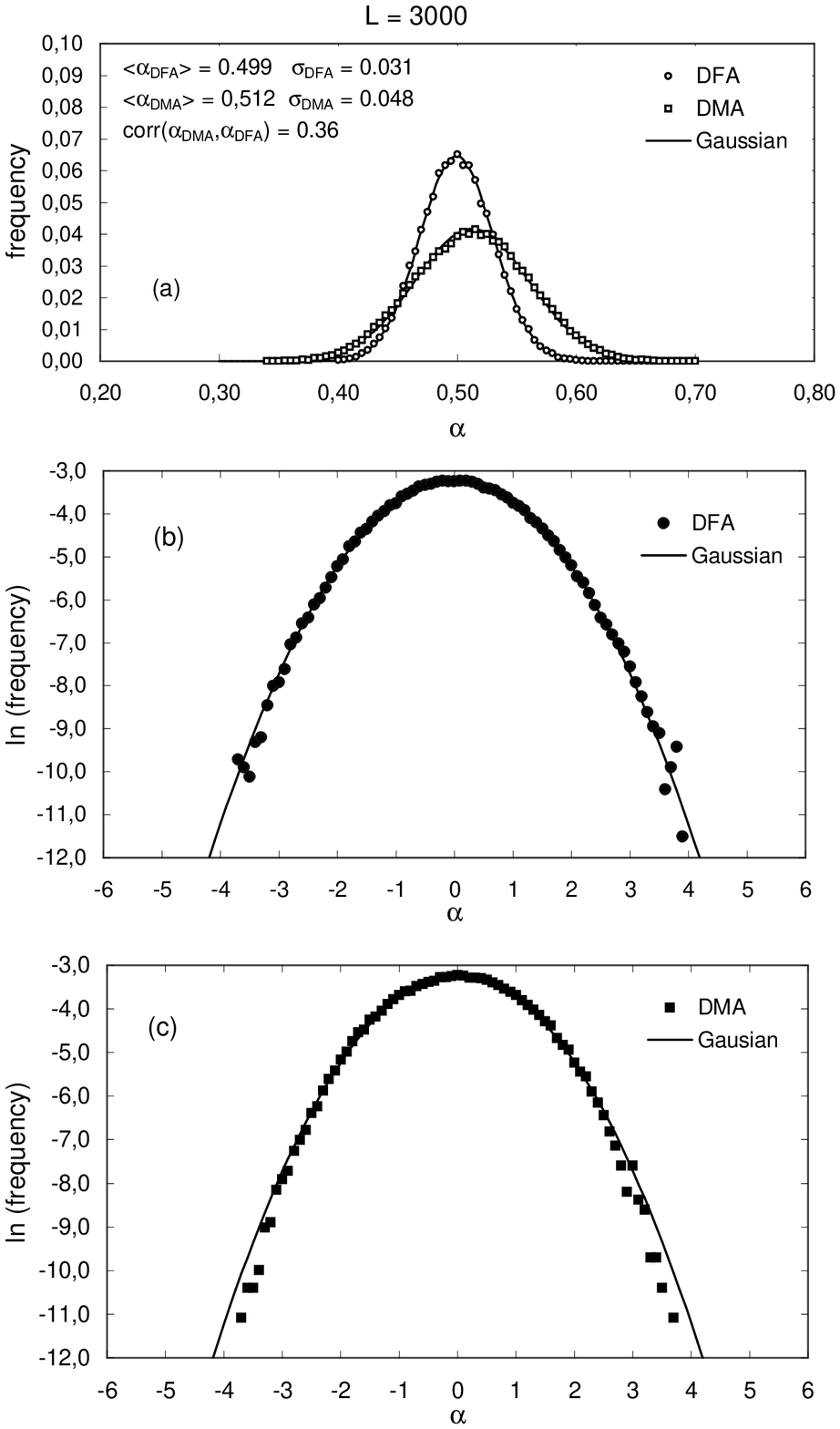,width=0.7\textwidth,angle=0}}
%\\
\caption{(a)Distribution of $\alpha$ exponents for series with
$L=3000$.
  (b)(c) Corresponding plots in semi-log scale.  }
\end{center}
\end{figure}

\begin{figure}
\begin{center}
{\epsfig{file=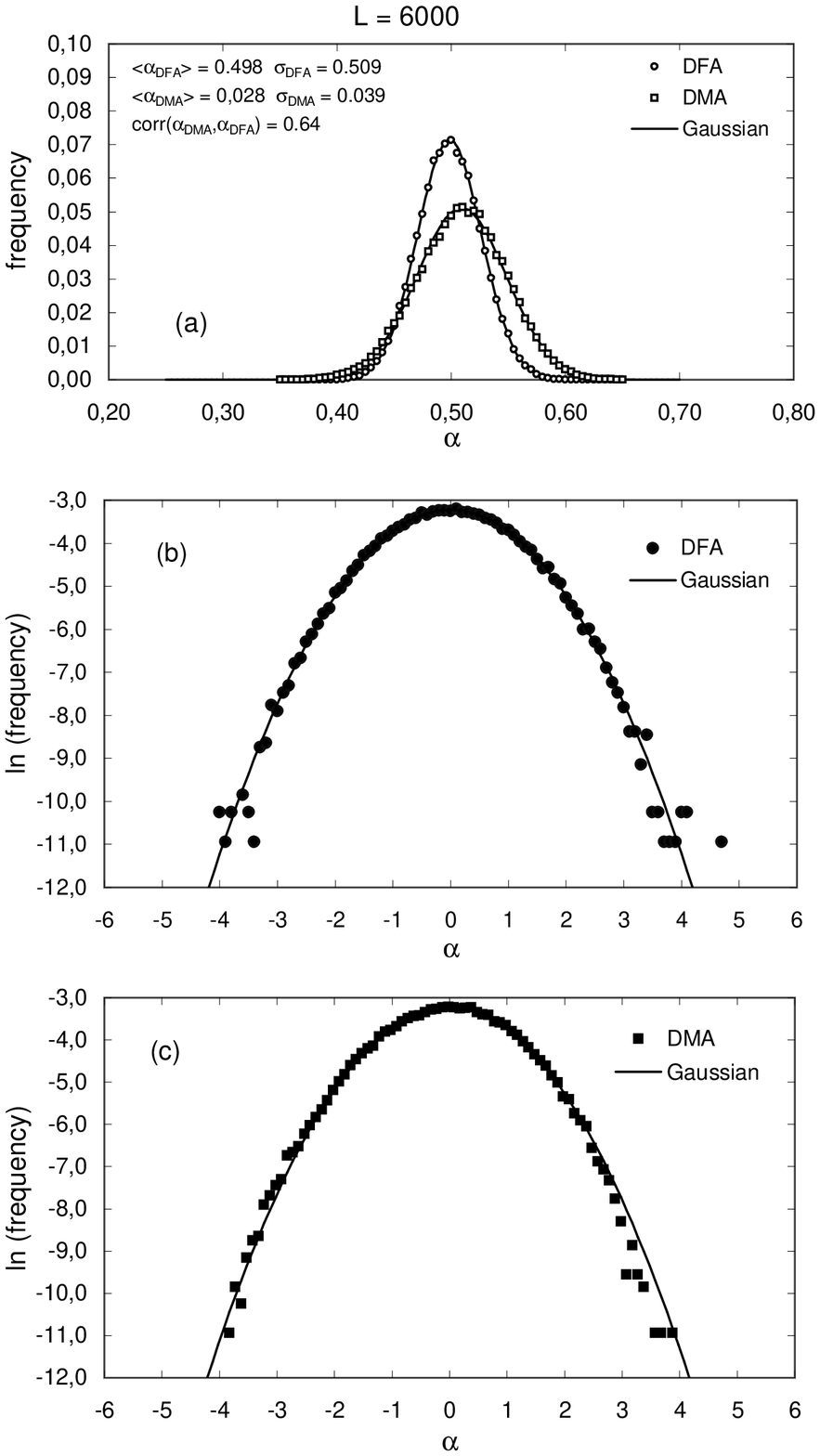,width=0.7\textwidth,angle=0}}
%\\
\caption{(a)Distribution of $\alpha$ exponents for series with
$L=6000$.
  (b)(c) Corresponding plots in semi-log scale.  }
\end{center}
\end{figure}

\begin{figure}
\begin{center}
{\epsfig{file=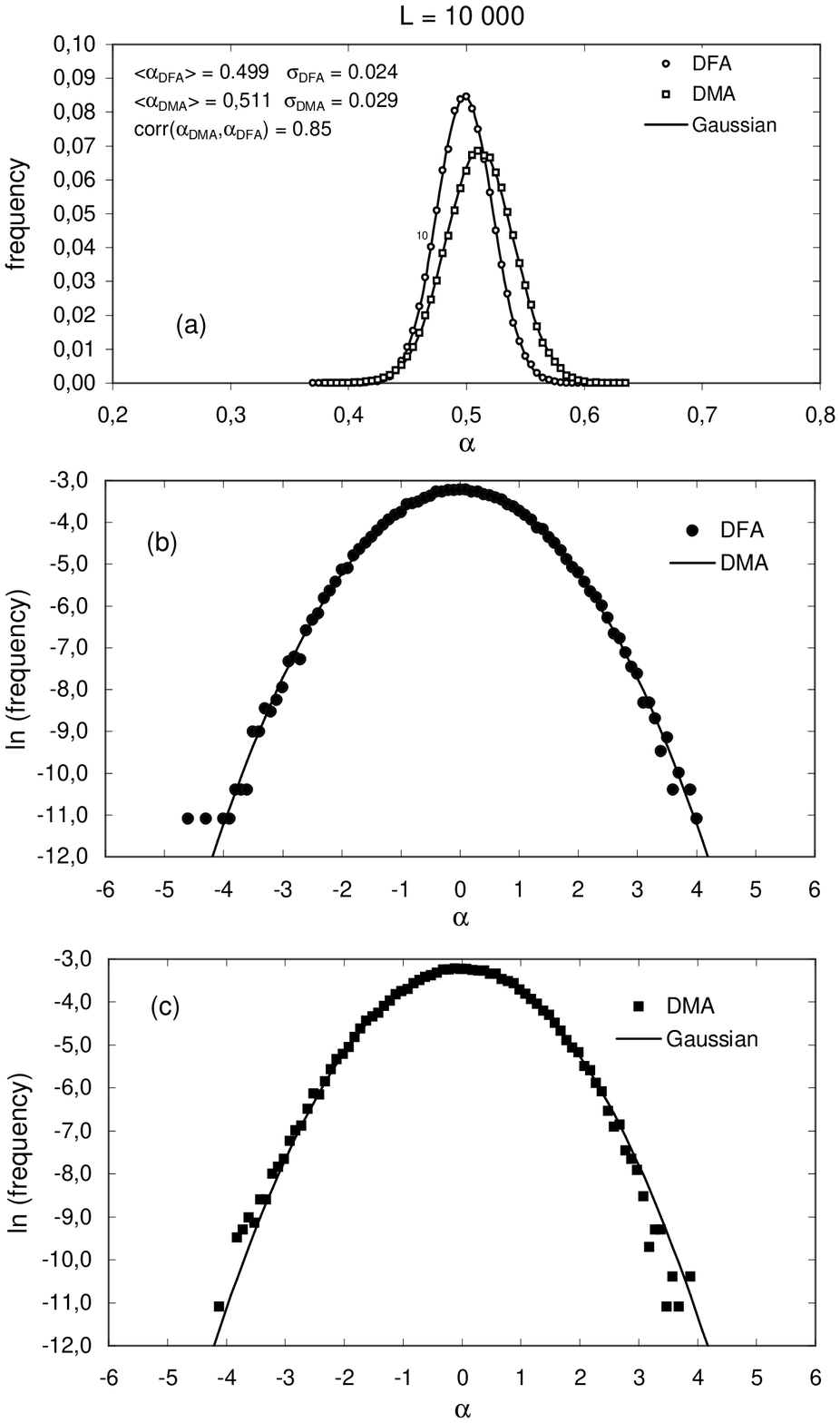,width=0.7\textwidth,angle=0}}
%\\
\caption{(a)Distribution of $\alpha$ exponents for series with
$L=10000$.
  (b)(c) Corresponding plots in semi-log scale.  }
\end{center}
\end{figure}

\begin{figure}
\begin{center}
{\epsfig{file=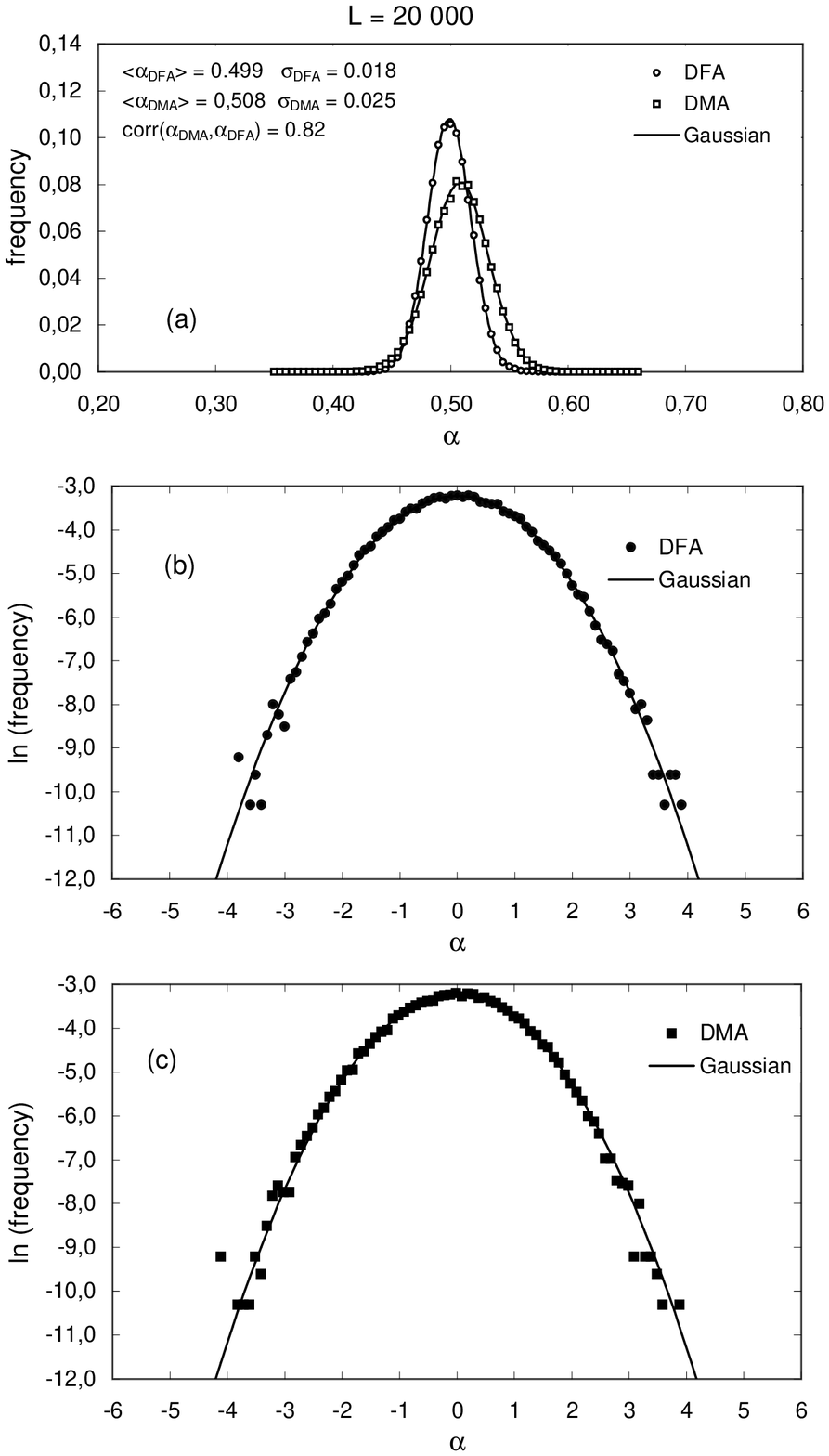,width=0.7\textwidth,angle=0}}
%\\
\caption{(a)Distribution of $\alpha$ exponents for series with
$L=20 000$.
  (b)(c) Corresponding plots in semi-log scale.  }
\end{center}
\end{figure}

\begin{figure}
\begin{center}
{\epsfig{file=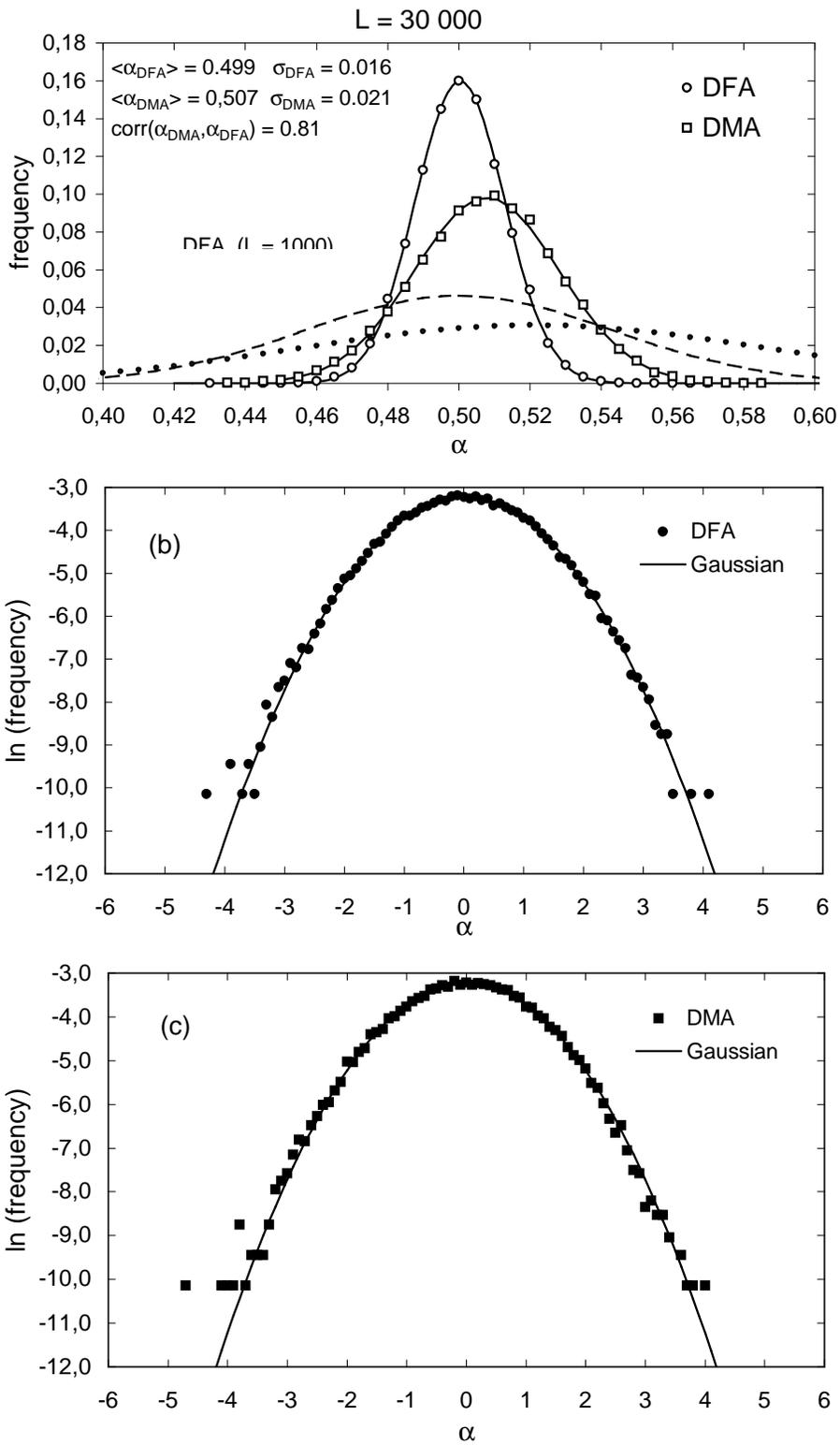,width=0.7\textwidth,angle=0}}
%\\
\caption{(a) Distribution of $\alpha$ exponents for series with
$L=30 000$.
   Additional lines represent $L=1000$ normal fit drawn for comparison in the same scale.
   (b)(c) Corresponding plots in semi-log scale.}
\end{center}
\end{figure}

\begin{figure}
\begin{center}
{\epsfig{file=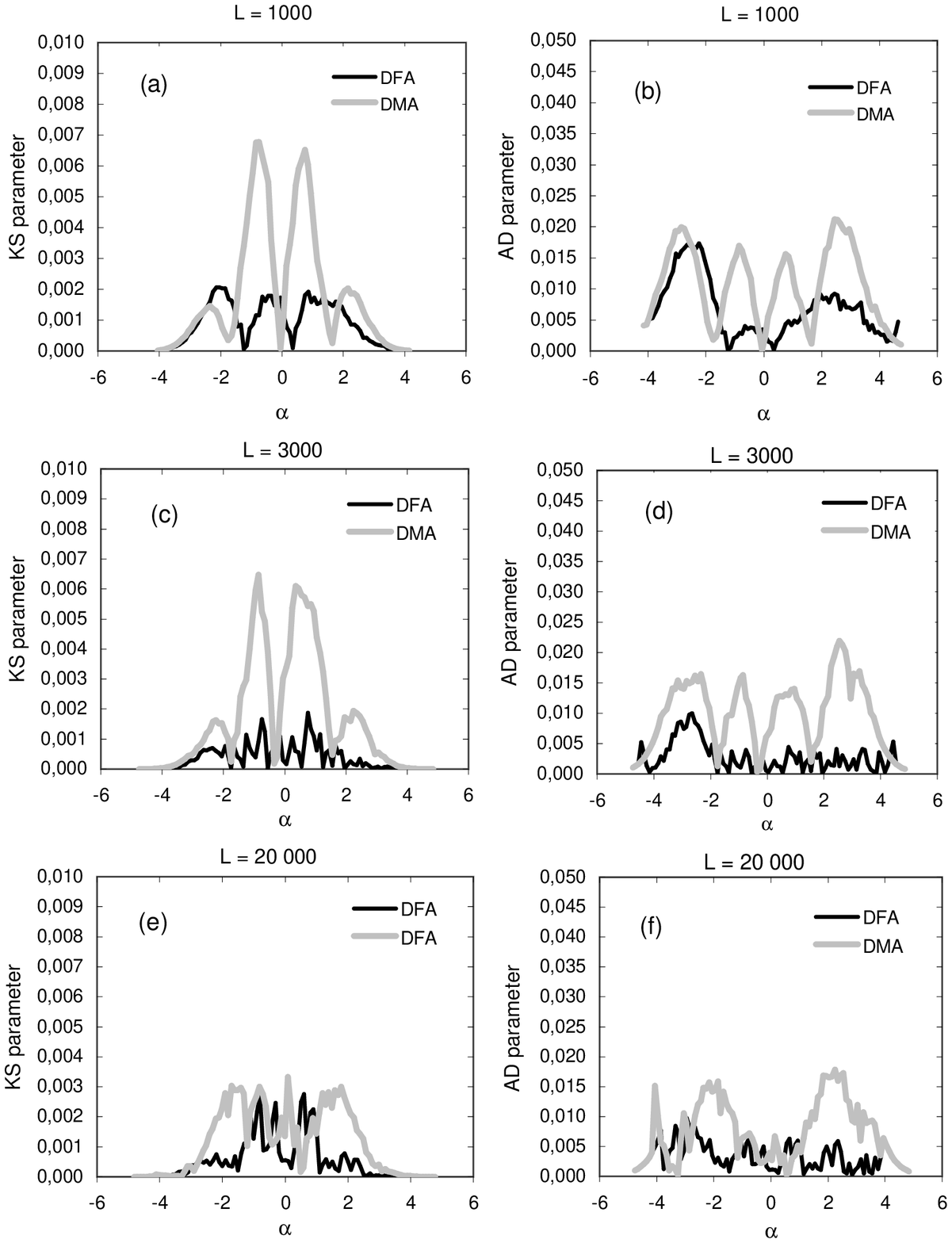,width=0.97\textwidth,angle=0}}
%\\
\caption{Kolmogorov-Smirnov (a)(c)(e) and Anderson-Darling (b)(d)(f) tests of
  correspondence between obtained distributions and the Gaussian one drawn respectively for time series
  of length $L=1000, L=3000, L= 20 000$}
\end{center}
\end{figure}
%%%%%%%%%???????????????

For this purpose we took samples of arithmetic Brownian time
series of length $L$ in the range $10^2-10^5$. Each sample
contained $N \sim 65 000$ series of fixed length. We tried to
cover uniformly the whole range of $L$ in log-scale keeping $L\sim
L_0 q^n$ with the approximate log step $q\sim 7/4$ to create
variety of lengths.
%%%%%%%%%

%%%%%%%%%%%

For any sample of fixed length series the averaged scaling range
$\langle \tau\rangle$ or $\langle \lambda\rangle$ has been
calculated for defined number of candidates ($\sim 30$) and the
corresponding standard deviation $\sigma_\tau$ ($\sigma_\lambda$).
The scaling range was taken as the range of $\tau$ or $\lambda$
variables strictly obeying scaling laws of
Eqs.~(\ref{gr7}),(\ref{gr11}) and assumed to terminate
respectively
 at
$\langle \tau\rangle-\sigma_\tau$ for DFA and $\langle
\lambda\rangle-\sigma_\lambda$ for DMA. Only series with
regression statistical correlation coefficient $R^2>0.98$ were
taken into account for $\alpha$ exponent extraction. For any
sample of time series a statistical distribution of $\alpha_{DFA}$
and $\alpha_{DMA}$ frequencies has been built.

%%%\\
%%%%%%%

%%%%
%%%%%%%%%%%%%

The full range of obtained distribution results is shown in
Fig.2-9. The first observation one makes is that for any length
$L$ both distributions fit very well normal distributions, but
with different parameters for the gaussian curve. We made all
plots also for centered and normalized $\alpha$ frequencies in
semi-log scale (Fig.2(b,c)--9(b,c)). Only small deviations from
the normal distribution are observed in tails - basically due to
smaller statistics there. A good correspondence with gaussian
curve is confirmed also in Kolmogorov and Anderson-Darling tests,
whose results are displayed in Table 1 and shown for chosen
lengths $L$ in Fig.10.

%%%%%% nowe miejsce dla tabelki
\bigskip

%%%% tabelka poczatek
\begin{table}[hb]
\caption{Kolmogorov-Smirnov (KS) and Anderson-Darling (AD) test
results for distribution of fluctuations in $\alpha$ exponent as a
function of  the method (DFA, DMA) and the length  $L$ of
time-series.}
\medskip
\begin{tabular}{|c|c|c|c|c|c|c|c|c|}
  \hline
 \begin{picture}(18,20)
\put(-15,20){\line(2,-1){48}} \put(12,10){L}
\end{picture}
% L
& 600  & 1000 & 1800 & 3000 & 6000 & 10 000 & 20 000 & 30 000 \\
  \hline
  KS$_{DFA}$
  & 5.0$\times$10$^{-3}$
  & 2.0$\times$10$^{-3}$
  & 2.0 $\times$10$^{-3}$
  & 2.0 $\times$10$^{-3}$
  &  2.4$\times$10$^{-3}$
   & 1.4$\times$10$^{-3}$
   & 2.7$\times$10$^{-3}$
   & 2.3$\times$10$^{-3}$
   \\ \hline
    KS$_{DMA}$
    & 6.4$\times$10$^{-3}$
    & 6.8$\times$10$^{-3}$
    &5.0$\times$10$^{-3}$
    & 6.5$\times$10$^{-3}$
      & 5.2$\times$10$^{-3}$
      &3.8$\times$10$^{-3}$
      & 3.3$\times$10$^{-3}$  & 3.6$\times$10$^{-3}$  \\
      \hline
  AD$_{DFA}$
  & 2.6$\times$10$^{-2}$ & 1.6$\times$10$^{-2}$
  & 1.0$\times$10$^{-2}$ & 1.0$\times$10$^{-2}$ & 1.3$\times$10$^{-2}$ & 8.0$\times$10$^{-3}$ & 9.7$\times$10$^{-3}$
  & 13.3$\times$10$^{-3}$  \\ \hline
  AD$_{DMA}$ & 3.6$\times$10$^{-2}$ & 2.1$\times$10$^{-2}$ & 1.8$\times$10$^{-2}$ & 2.1$\times$10$^{-2}$ & 2.7$\times$10$^{-2}$ & 2.1$\times$10$^{-2}$ &
  1.8$\times$10$^{-2}$ & 2.5$\times$10$^{-2}$ \\
  \hline
\end{tabular}
\end{table}
%%%% tabelka koniec
%%%%%%%

One may notice that the standard deviation $\sigma_{DFA}$ of
$\alpha_{DFA}$ scaling parameters is always smaller than the
corresponding standard deviation $\sigma_{DMA}$ of $\alpha_{DMA}$
exponents, and both standard deviations decrease when $L$ grows.
This can be explained in terms of different sensitivity of DFA and
DMA techniques to the presence of random autocorrelations in time
series. Such autocorrelations are naturally randomly distributed
in any sample of generated time series and hence a distribution of
$\alpha$ exponent is normal. The probability of random
autocorrelations is bigger for short time series, where all
statistical fluctuations manifest in a more vivid way. When $L$
increases, their influence on the presumed global autocorrelation
in series can be neglected. Therefore, both standard deviations
$\sigma_{DFA}$ and $\sigma_{DMA}$ drop with increasing $L$.
However, we always observe $\sigma_{DFA}<\sigma_{DMA}$, what
indicates that DMA technique is more sensitive to such
"autocorrelation noise" than DFA one.

One may look at this problem also from another side - like in
Fig.~11. Here we have drawn several plots of DFA and DMA analysis,
i.e. $\ln F$ vs $\ln \tau$ or $\ln \lambda$  for several
corresponding artificial Brownian series of length $L=1000$. It is
seen that deviations from the strict power law behavior, if occur,
are more drastic for DMA than for DFA case and the dispersion of
produced slopes is also larger for DMA than for DFA, despite the
fact that DMA plots are more smooth in comparison with DFA ones.
%%%%%%%%%%%%%% 666666666666666

The next observation concerns the mean values. One gets
%\sloppy
$\langle \alpha_{DFA}\rangle_N<\langle \alpha_{DMA}\rangle_N$
 for
all $L$, where $\langle.\rangle_N$ is taken over a sample of $N$
time series. A~clear shift of the central DMA values to the right
with respect to DFA ones (see Figs.~2--9(a)) does not suggest
however the presence of systematic relation between $\alpha_{DMA}$
and $\alpha_{DFA}$. Indeed evaluating the correlation coefficient
(values are shown in the description of Fig. 2--9(a)):

\begin{eqnarray}
\label{gr13} corr  %%%%%%%%%%%????????????????????/
(\alpha_{DFA}, \alpha_{DMA}) =
\frac{\langle\alpha_{DFA}\alpha_{DMA}\rangle_N -
\langle\alpha_{DFA}\rangle_N\langle\alpha_{DMA}\rangle_N}{\sigma_{DFA}\sigma_{DMA}}
\end{eqnarray}
one finds it increasing with $L$, but it never indicates the full
correlation. Its value  is maximal for large $L$, where $
corr(\alpha_{DFA},\alpha_{DMA})\sim 0.8$ for $L\sim 10^4 - 10^5$.

{This situation is  graphically illustrated in Fig.~12, where a
correlation plot $\alpha_{DFA}$ vs $\alpha_{DMA}$ is shown for
Hurst exponent values obtained for $L=3000$, $L=10 000$ and $L=30
000$ series. From the asymmetry of plots against diagonal one
notices that DMA gives higher values than DFA method in most
series. This result is independent on the length of time series.
In fact the percentage excess of cases $n_{+}$, where
$\alpha_{DMA}
> \alpha_{DFA}$ over the cases where
 $ \alpha_{DMA}< \alpha_{DFA}$ ($n_{-}$), i.e.:}

%%%%%%%%%%%%%%%%%%%%%%%%

%%%%%%%%%%%%%
\begin{figure}
\begin{center}
{\epsfig{file=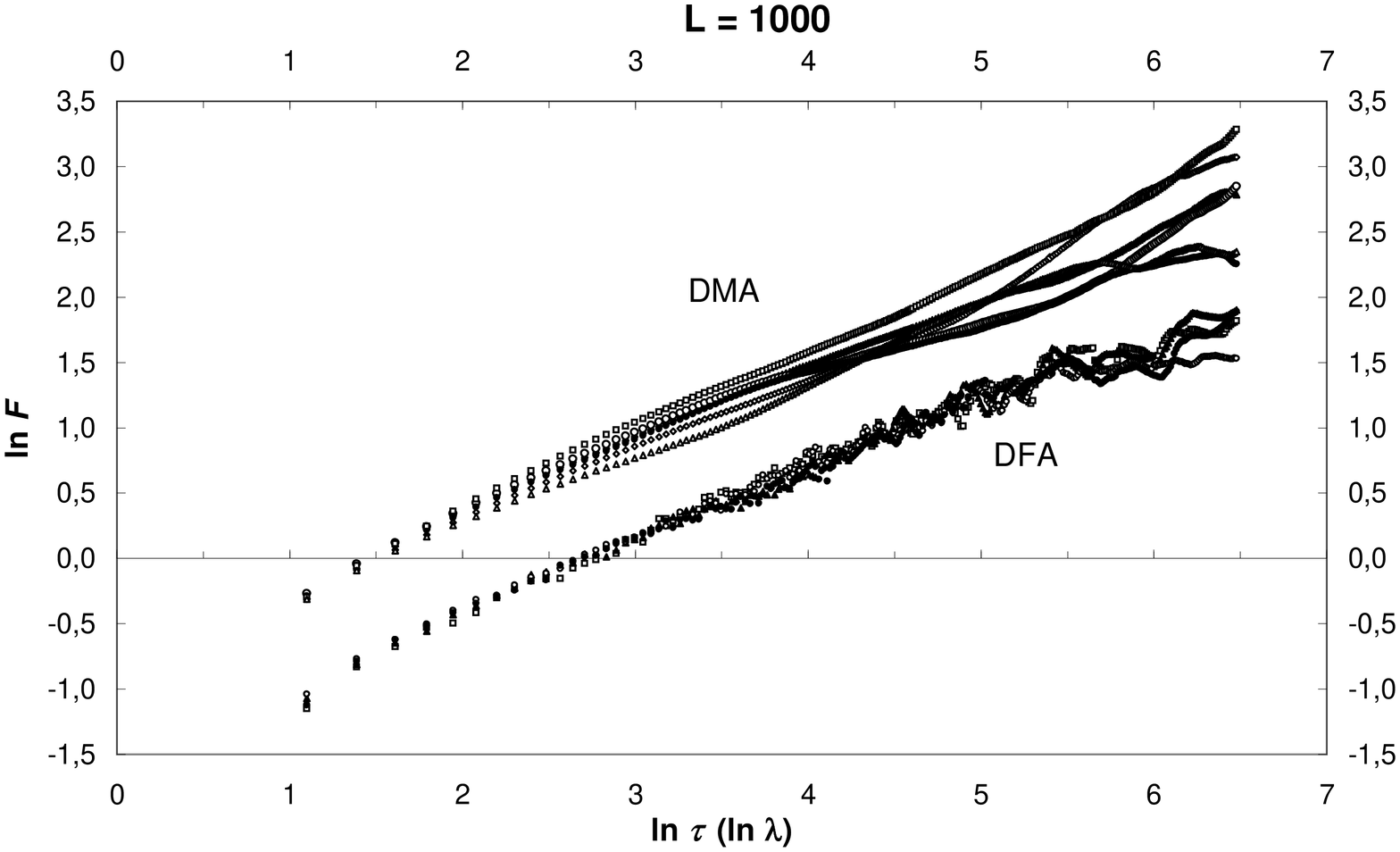,width=0.4\textwidth,angle=-90}}
%\\
\caption{Examples of DMA and corresponding DFA plots $\ln F$ vs
$\ln \tau (\ln \lambda)$ for several randomly  chosen Brownian
integer time series of length $L=1000$. }
\end{center}
\end{figure}
%%%%%%%%%???????????????

\begin{figure}
\begin{center}
{\epsfig{file=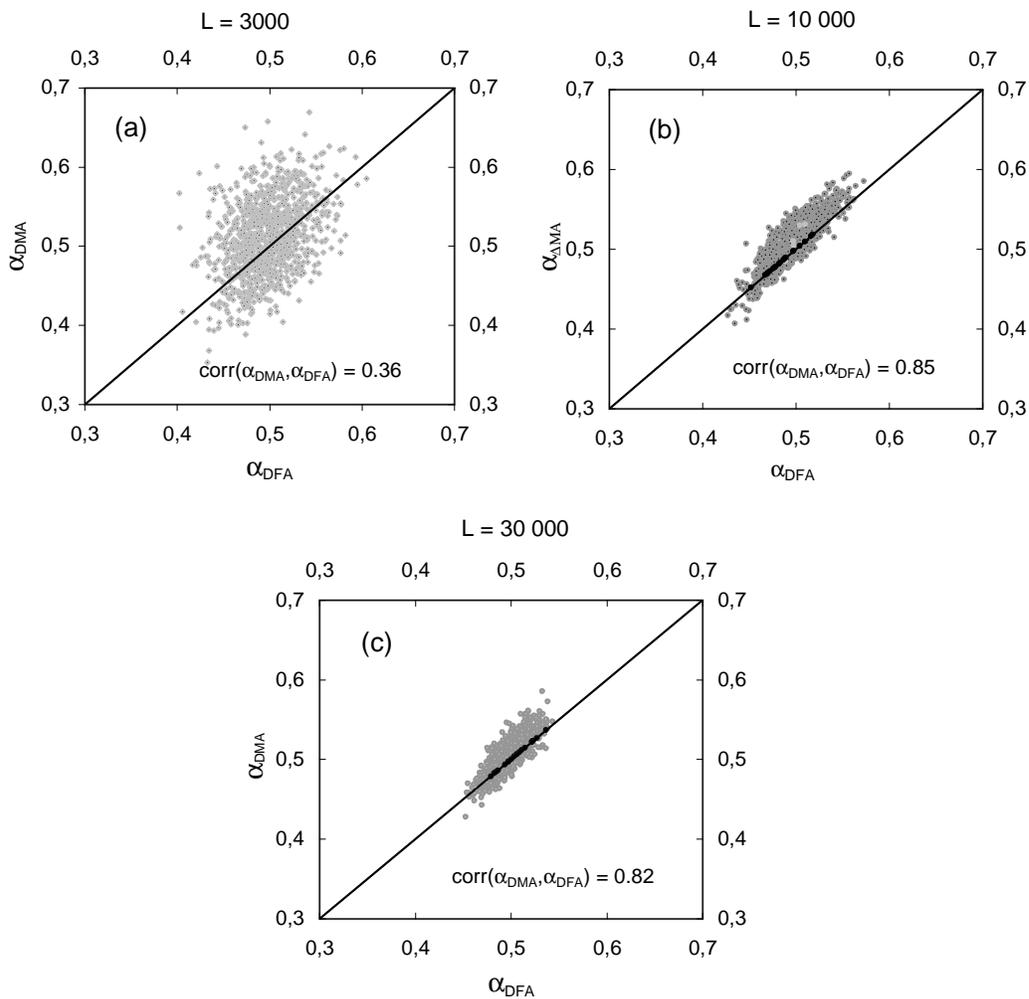,width=0.8\textwidth,angle=0}}
%\\
\caption{Correlation plot $\alpha_{DFA}$ vs $\alpha_{DMA}$ for the
sample of $65 000$ Brownian time series of length a) $L= 3 000$,
 b) $L= 10 000$, c) $L=30 000$. }
\end{center}
\end{figure}
%%%%%%%%%???????????????

\begin{figure}
\begin{center}
{\epsfig{file=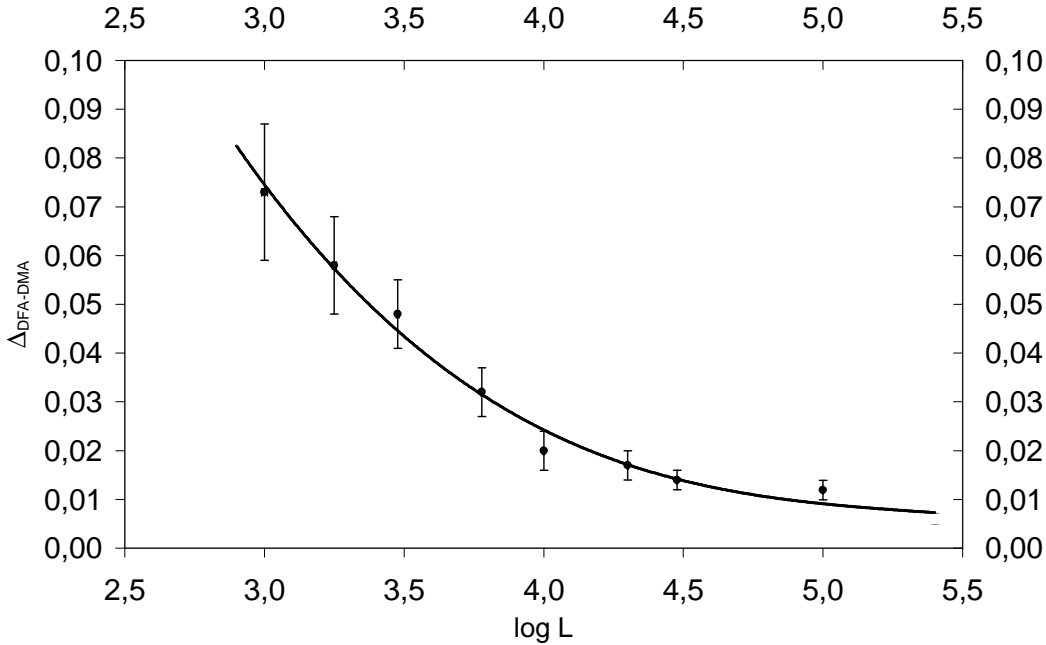,width=0.5\textwidth,angle=-90}}
%\\
\caption{A mean difference $\Delta_{DFA-DMA}$ between
 $\alpha_{DFA}$ and  $\alpha_{DMA}$ exponents calculated for the same series
 as a function of the series length $L$. Marked error bars $\delta \ \Delta_{DFA-DMA} \sim \delta
 \alpha_{DFA}+ \delta\alpha_{DMA} $ correspond to uncertainties in slope determination
  in regression analysis for both methods.}
\end{center}
\end{figure}
%%%%%%%%%???????????????

%%%%
%%%%%%??????????????????
\begin{eqnarray}
\label{gr14} \delta_{\pm} = \frac{n_{+}-n_{-}}{n_{+}+n_{-}}
\end{eqnarray}
changes from $\sim 20\%-25\%$ for series with $L<10 000$ up to
$\sim50\%$ for longer series.
 %%%%%%%%%%%%%%%%%

%%%%

It is obvious therefore that the mean of difference
$\delta_{DFA-DMA}$, where
\begin{eqnarray}
\label{gr15} \delta_{DFA-DMA} = \langle\alpha_{DFA} -
\alpha_{DMA}\rangle_N
\end{eqnarray}
is not a good measure of 'distance' between two investigated
methods. It is more convenient to define this distance in a
standard way, i.e.:
\begin{eqnarray}
\label{gr16} \Delta_{DFA-DMA} = \left({\langle(\alpha_{DFA} -
\alpha_{DMA})^2\rangle_N}\right)^{1/2}
\end{eqnarray}

%%%%%%%%%%%%%%%
The sufficient number of time series samples of various length has
been worked out to find a relationship $\Delta_{DFA-DMA}(L)$. The
polynomial best fit for the collected data is drawn in Fig.~13
with error bars coming from the uncertainties in slope
determination. This plot indicates that the average displacement
between $\alpha_{DFA}$ and $\alpha_{DMA}$ exponents for a given
time series ranges from $15\%$ for series with $L\leq10^3$, down
to $2\%$ for long series ($L\sim10^5$). The latter value is much
smaller than one reported in \cite{bgr5}. The fastest drop in
DFA-DMA distance is observed for medium length series, i.e. when
$L\sim 10^3-10^4$. For such series $\Delta_{DFA-DMA}$ makes on the
average $\sim 10\%$ of $\alpha_{DFA}$ value.
%%%%%%%%%%%%%9999999999999999

%%%%%%%%%%%%%%%%%%

This might be of interest if more detailed study of $\alpha$
exponent is required for more exact predictions to be made(e.g.
heart diseases, finances, etc.). The plot in Fig.~13 may also
suggests that $\Delta_{DFA-DMA}\rightarrow 0$ when
$L\rightarrow\infty$. The latter case has not been explored in
details.
%%%%%%%%%%%%%

\section{Conclusions}
We report from the analysis of artificial Brownian integer time
series and from the collected data that, on the average, DMA
method overestimates  Hurst exponent values in comparison with DFA
technique. This result contradicts to some previous hypothesis in
literature. The DMA method seems to be also more sensitive to the
presence of random fluctuations in autocorrelations in time series
than DFA analysis does. In many practical situations, especially
for shorter series, it might be a disadvantage leading to the
false signal of not really existing, global
autocorrelations in time series.\\
The mean distance between two methods, i.e. the mean difference
between $\alpha_{DFA}$ and $\alpha_{DMA}$ exponents calculated for
the time series of given length $L$ is a decreasing function of
$L$. For shorter series ($L\leq 6 000)$ this distance reaches
$\sim 15\%$ what might be important in precise
determination of $\alpha$ exponent for such series.\\
 There are
some open questions. It is not exactly clear where the scaling law
exactly starts or terminates, so one needs a more strict
requirements how the scaling range should be determined for DFA
and DMA techniques and how uncertainties in the choice of scaling
range are related to uncertainties in the scaling exponent
$\alpha$. This work is now in progress \cite{bgr14}.

\end{document}